\documentclass[11pt,a4paper]{article}
\usepackage{ifpdf}
\usepackage{times}
\usepackage[scale={0.8,0.9},centering,includeheadfoot]{geometry}
\usepackage{enumitem}
\usepackage{multicol}
\usepackage{moreverb,url}
\usepackage{graphicx}
\usepackage{subfig}
\usepackage[colorlinks,bookmarksopen,bookmarksnumbered,citecolor=red,urlcolor=red]{hyperref}
\usepackage{listings}
\usepackage{authblk}
\usepackage{amsmath, xparse}
\usepackage{verbatim}

\newtheorem{recommendation}{Recommendation}

\def\qedbox{\ifvmode\else\unskip\fi~\penalty10000
    \hfill{\large$\blacksquare$}}

\def\Fig#1{Figure~\ref{#1}}

\def\Sec#1{Section~\ref{#1}}

\def\eref#1{(\ref{#1})}

\def\mm#1{\ensuremath{\boldsymbol{#1}}} 
\def\Var{\text{Var}}

\title{A note on intrinsic Conditional Autoregressive models for
    disconnected graphs}

\author{Anna Freni-Sterrantino\footnote{Corresponding author, Small Area Health Statistics Unit, Department
    of Epidemiology and Biostatistics,
    Imperial College London, United Kingdom\\
Email:a.freni-sterrantino@imperial.ac.uk}\\
\and 
Massimo Ventrucci\footnote{Department of Statistics, University of Bologna, Bologna, Italy}\\
\and 
H{\aa}vard Rue \footnote{CEMSE Division, King Abdullah University of Science
    and Technology, Saudi Arabia}}

\date{}

\begin{document}

\maketitle

\begin{abstract}
    In this note we discuss (Gaussian) intrinsic conditional
    autoregressive (CAR) models for disconnected graphs, with the aim
    of providing practical guidelines for how these models should be
    defined, scaled and implemented. We show how these suggestions can
    be implemented in two examples on disease mapping.
\end{abstract}



\section{Introduction}
\label{sec:intro}

Conditional Autoregressive (CAR) models are widely used to represent
local dependency between random variables. They are numerous
applications in disease mapping \cite{art430, book116} and imaging
\cite{art149}. In this paper, we introduce the specification of a CAR
model on a disconnected graph and then show application on two disease
mapping examples. To avoid unnecessary technicalities, we will
throughout this note, only discuss a simple case, leaving the straightforward generalisation to the reader.

Disease mapping concerns the study of disease risk over a map of
geographical regions. Let assume the study area is a lattice of
$i=1,...,n$ non overlapping regions and $y_i$ is the number of cases
of a given disease in region $i$. For a rare disease, a Poisson model
is assumed, $y_i | \theta_i \sim \text{Po}(\theta_i)$, $i=1,...,n$,
with mean $\theta_i = E_i r_i$, where $E_i$ is the expected
cases count for the disease under study (computed using the disease
rates and demographic characteristics of a reference population) and
$r_i$ is the relative risk, such that $r_i>1$ ($r_i<1$) means higher
(lower) risk associated with living in region $i$, while $r_i$ close to 1, indicates 
little difference between observed and expected in the $i-th$ region.
 The relative risk can be modelled in terms of the effect of a
covariate $z$, e.g.\ pollution, as
$\log(r_i)=\alpha + \beta z_i + x_i$, where $\alpha$ and $\beta$ are
respectively the baseline log risk and the effect of pollution. Value
$x_i$ is a random effect capturing extra Poisson variability possibly
due to unobserved risk factors.

When residual variability is spatially structured, a popular approach
is to model the random effects with an intrinsic CAR, i.e.
$x_i | x_{-i}, \kappa \sim \mathcal{N}(\sum_{j:i \sim j} x_j/n_i, (n_i
k)^{-1} )$, $i=1,...,n$. The precision hyper-parameter $\kappa$ regulates
the degree to which $x_i$ is shrunk to the local mean
$\sum_{j:i \sim j} x_j/n_i$, which is the average of the random
effects over its $n_i$ neighbours $j:i \sim j$. This model is
intrinsic in the sense that the overall mean is left unspecified and
can be identified only when adding a linear constraint, such as
$\sum_i x_i=0$.

On the applied side, this model is useful, for instance, when
underlying population at risk is heterogeneous over the study area,
with sometimes small expected counts in small regions. In these cases,
the standardised incidence/mortality ratios $y_i/E_i$ are affected by
large variances, hence a map of those ratios give a noisy
representation of the disease risk over the study area. A CAR prior
for the region-specific random effects $x_i$'s allows borrowing
strength of information between neighbours, yielding a more reliable
smooth map for the disease risk.

The definition of a CAR model starts by specifying a graph. A graph is
a collection of nodes and edges representing, respectively, regions
and neighbouring relationships between them. A graph is connected if
there is a path (i.e. a set of contiguous edges) that connects each
node to at least another node. Within a connected graph, specification
of neighbouring relationships is clear (each node has at least one
neighbour) and definition of a CAR model follows straightforwardly.

The specification of a CAR model on a disconnected graph is undefined and how should be carried out. 
There are essentially two types of disconnected graphs:
first, a graph containing an island (a singleton node with no
neighbours), second, a graph split in different sub-graphs (each of
them being a connected graph).

In literature, there is a lack of attention~\cite{Knorr} on the
definition of a CAR for a disconnected graph, and/or on the properties
of a CAR when the graph is disconnected. The only reference on this
topic is Hodeges et al.~\cite{Hodges2003} who discuss the form of the
normalizing constant. One difficulty is how to deal with random effect
in a singleton. \texttt{GeoBUGS} manual~\cite{geobugs} offers some
guidelines on this, with a default option which is to set $x_i$ to
zero, if $i$ is a singleton. This practice is equivalent to enforce a
sum-to-zero constraint $x_i=0$ on the singleton random effect: back to
the disease mapping example, this automatically sets
$r_i=\exp(\alpha + \beta z_i)$, if the baseline/covariate component is
included in the model.

There are two issues with this approach. The first one is that it seems
too restrictive, in the sense that even though a singleton random effect $x_i$
cannot capture spatially structured variability because it has no
local mean to shrink to, $x_i$ should at least be allowed to model
unstructured variability, hence shrinking towards a global mean. The second one and more general issue, with CAR models, regards
scaling \cite{art521} which is important in order to interpret the
prior assigned to the hyper-parameter $\kappa$. Care is needed when scaling
the precision of a CAR model defined on a disconnected graph.

In the rest of this paper we discuss in detail the aforementioned two
issues. In particular, in Sections~\ref{sec:intrinsic-car} we define
the intrinsic CAR model. In Section ~\ref{sec:scale-connected} and
~\ref{sec:scale-disconnected} we provide recommendations on
appropriate scaling for the precision of a CAR model defined on
connected and disconnected graphs, respectively. In Section
\ref{sec:linear-constraints} we give recommendations on linear
constraints and discuss computation of the normalizing constant. In
Section~\ref{sec:examples} we illustrate the proposed methods in two
examples on disease mapping involving two different types of graphs.
We conclude with a discussion in Section~\ref{sec:summary}.

\section{Intrinsic CAR models}
\label{sec:intrinsic-car}

In its simple form, the density of an intrinsic CAR model for
$\mm{x} = (x_1, \ldots, x_n)^{T}$ is
\begin{equation}\label{eq2}%
    \pi(\mm{x} \mid \kappa) \propto \frac{1}{Z_n(\kappa)}
    \exp\left(
      -\frac{\kappa}{2} \sum_{i\sim j} (x_i - x_j)^{2}
    \right)
\end{equation}
where $i \sim j$ is the set of all pair of \emph{neighbours}, $\kappa$
is a precision parameter, and $Z_n(\kappa)$ is a normalising constant
that we will return to later on. What qualify as a ``neighbour'' is
application dependent and part of the model specification. For
example, in many spatial applications, two regions ($i$ and $j$, say)
are considered to be neighbours if they share a common border. The
interpretation of \eref{eq2} is that similarity between two neighbours
are encouraged, and this induce a smoothing effect between neighbours,
and thereby between neighbours of neighbours, and so on. We can
formalize this, by defining a undirected \emph{graph}
${\mathcal G} = ({\mathcal V}, {\mathcal E})$, with a set of vertices
${\mathcal V} = \{1, 2, \ldots, n\}$, and the edges ${\mathcal E}$ are
the (unordered) set of neighbours. We then say that the intrinsic CAR
is defined with respect to the graph ${\mathcal G}$.

The precision parameter $\kappa$ determines the amount of smoothing
and it is commonly estimated from data. The density \eref{eq2} is
improper or \emph{intrinsic}, in the sense that is invariant to adding
the same constant to all the $x_i$'s, as an example of a first order
polynomial intrinsic CAR. Weighted and higher order
polynomial (and non-polynomial) intrinsic CAR's can be defined
similarly; see \cite[Ch.~3]{book80} for a thorough discussion.

We will focus on discussing the simplest case of an intrinsic CAR
models for disconnected graph~\eref{eq2} to avoid technicalities which
are discretional to understand the ideas, that are easily generalized
to other types of intrinsic CAR models.

\section{Scaling of an intrinsic CAR defined for a connected graph}
\label{sec:scale-connected}

Intrinsic CAR models has an unresolved issue with \emph{scaling},
which is not immediate from studying~\eref{eq2}. The basic reference
is S{\o}rbye SH et al.\cite{art521}, which we will base our arguments
in this section.

\begin{figure}[!tbp]
    \centering
    \subfloat[]{\includegraphics[scale=.7,width=0.4\textwidth]{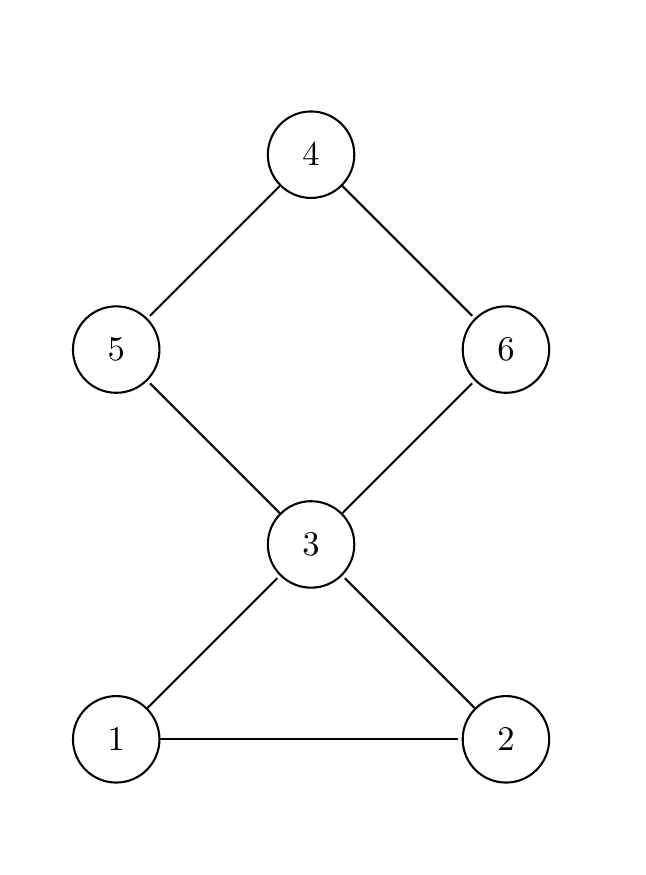}
        \label{fig:1}} \hfill
    \subfloat[]{\includegraphics[scale=.7,width=0.4\textwidth]{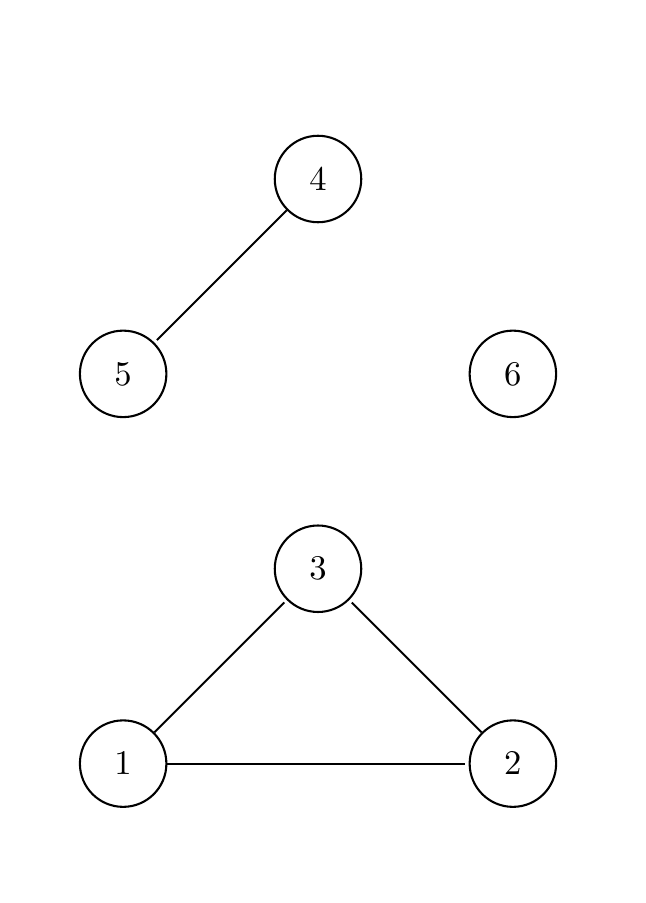}
        \label{fig:2}}
    \caption{The two graphs used in the discussion: (a) a connected
        graph, and (b) a disconnected graph.}
\end{figure}

We will assume that the graph ${\mathcal G}$ is connected, meaning
that there is a path between all pair of nodes in the graph. An
example of a connected graph is shown in \Fig{fig:1}.
The intrinsic CAR model~\eref{eq2} defined for this graph, has
precision matrix
\begin{equation}\label{eq1}%
    \mm{Q} = \kappa\begin{bmatrix}{}
        2 & -1 & -1 &  &  &  \\ 
        -1 & 2 & -1 &  &  &  \\ 
        -1 & -1 & 4 &  & -1 & -1 \\ 
        &  &  & 2 & -1 & -1 \\ 
        &  & -1 & -1 & 2 &  \\ 
        &  & -1 & -1 &  & 2 \\ 
    \end{bmatrix}.
\end{equation}
The zeros are not shown. The intrinsic density is invariant for adding
a constant to $\mm{x}$, meaning that $\mm{x}$ and $\mm{x} +c \mm{1}$
has the same improper density. However, what is of practical interest
here, is \emph{how} and \emph{how much} this model varies around its
mean value $\overline{\mm{x}}$, i.e.\ how $\mm{x}$ vary if we impose a
sum-to-zero constraint $\mm{1}^{T}\mm{x} = 0$. The critical issue is
that this is a complicated function of the graph ${\mathcal G}$, for
which we have no good intuition. For our example the (conditional on $\kappa$) marginal variances are $0.53/\kappa$, $0.53/\kappa$, $0.19/\kappa$,
$0.53/\kappa$, $0.44/\kappa$ and $0.44/\kappa$ for $x_1, \ldots, x_6$,
which we interpret as follows.
\begin{enumerate}
\item The issue - that (conditional) marginal variances are different -
    is a \emph{feature} of the intrinsic CAR model, and a consequence
    of that the conditional variance, $\Var(x_i | \mm{x}_{-i})$, is
    inverse proportional to the number of neighbours of node $i$
    (Eq.3.32\cite{book80}).
\item The `typical marginal variance' is confounded with our
    interpretation of the precision parameter $\kappa$, and we do not
    know a-priori what $\kappa=1$ means in terms of a typical marginal
    variance. In the Bayesian framework, this is crucial issue, since
    we need to impose a prior distribution for $\kappa$. We need to
    address what $\kappa$ means in terms of its impact on the model.
\end{enumerate}

The solution out of this appearently dilemma, is simply to scale
$\mm{Q}$ so that the typical marginal variance is 1 when $\kappa=1$.
S{\o}rbye and Rue\cite{art521} recommend to use the geometric mean,
which in our example, gives the following scaled precision matrix

\begin{equation}\label{eq3}%
    \mm{Q}_{\text{scaled}} = \kappa c 
    \begin{bmatrix}{}
        2 & -1 & -1 &  &  &  \\
        -1 & 2 & -1 &  &  &  \\
        -1 & -1 & 4 &  & -1 & -1 \\
        &  &  & 2 & -1 & -1 \\
        &  & -1 & -1 & 2 &  \\
        &  & -1 & -1 &  & 2 \\
    \end{bmatrix}
\end{equation}    
where $c = 0.4219\ldots$. The most important consequence of this
scaling, is that $\kappa$ is now the typical precision and not only
\emph{a precision parameter}. This makes it possible to define a
meaningful prior distribution and a clear interpretation for $\kappa$.
There is a long tradition to prefer model parameters with a good and
clear interpretation.

Our recommandation is clear and unambiguous.
\begin{recommendation}
    We recommend to scale intrinsic CAR models defined with regard to
    connected graphs.
\end{recommendation}

The scaling parameter $c$ can be computed as the geometric mean of the
diagonal of the generalized inverse of \mm{Q} when $\kappa=1$;
hereafter we denote $\mm{Q}$ as $\mm{R}$, if $\kappa=1$. However, this
is not a computational efficient way to compute it has its an
${\mathcal O}(n^{3})$ operation. A better approach, is to make use of
the graph of the model and the knowledge of the null-space, and treat
\mm{R} as a sparse matrix; see Rue and Held \cite[Ch.~2.4]{book80} for
background and details. The scaling $c$ can then be computed from a
rank one correction of the marginal variances from the unconstrained
model, see Rue et al.\cite{art375} for technical details about the
recursions leading to the marginal variances. The computational cost
will be be ${\mathcal O}(n^{3/2})$ for typical spatial graphs, which
is a huge improvement. The \texttt{R}-function
\texttt{inla.scale.model()} in the \texttt{R-INLA} package (see
\texttt{www.r-inla.org}) is an efficient implementation of this.

\section{Scaling of an intrinsic CAR defined for a disconnected graph}
\label{sec:scale-disconnected}

A graph is disconnected if it is not connected. The practical
interpretation of this related to~\eref{eq2}, is that there are
``islands'' in the graph or nodes with no neighbours; we denote these
nodes as singletons. Such cases easily appear if the graph is
contructed from a map where we can have islands, or regions that are
physically disconnected from the rest of the area. \Fig{fig:2} shows a
disconnected graph with three \emph{connected components} of size $3$,
$2$ and $1$. We will use this graph as reference, in this section.

A direct application of the intrinsic CAR model for graph in
\Fig{fig:1}, gives the precision matrix
\begin{equation}\label{eq4}%
    \mm{Q} = \kappa
    \begin{bmatrix}{}
        2 & -1 & -1 &   &   &   \\
        -1 & 2 & -1 &   &   &   \\
        -1 & -1 & 2 &   &   &   \\
        &   &   & 1 & -1 &   \\
        &   &   & -1 & 1 &   \\
        &   &   &   &   &  0 \\
    \end{bmatrix}.
\end{equation}
This matrix is singular with rank-deficiency of $3$, since the density
is invariant when we add a constant to each connected component. There
are several unfortunate issues with this intrinsic CAR model, simply
because the \emph{implicite} assumption behind~\eref{eq2} is that the
graph is connected. Additional to the issues discussed in
Section~\ref{sec:scale-connected}, we have the following due to the
disconnected graph.
\begin{itemize}
\item Node $6$ has no neighbours so $Q_{6,6} = 0$ and we have a
    constant density for $x_6$. This can lead to an improper posterior
    distribution.
    To see this, consider the following model for an observed count,
    $y_6$, in node $6$,
    \begin{eqnarray}
      y_6|x_6 & \sim & \text{Po}(\exp(x_6)) \nonumber\\
      \pi(x_6|\kappa) & \propto & const \label{eq2:constant-prior}
    \end{eqnarray}
    where $x_6=\log(\theta_6)$ is the Poisson mean, $\theta_6$, in
    logarithmic scale. The constant prior \eref{eq2:constant-prior}
    implies an improper prior on the Poisson mean, i.e.
    $\pi(\theta_6) \propto 1/\theta_6$. If a zero count is the case,
    then $\pi(\theta_6 | y_6=0)\propto \exp(-\theta_6) / \theta_6$,
    which is improper. In other words, the constant prior for the
    singleton makes it difficult for the singleton random effect to
    shrink to the global mean. Also, this goes against the purpose of
    using~\eref{eq2} in the first place, which is to do smoothing and
    borrowing strength.
\item The connected components for $(x_1, x_2, x_3)$ and $(x_4, x_5)$
    are defined as on a connected graph. Even though this is
    reasonable within each connected components, it is not reasonable
    when we compare across connected components. As discussed in
    Section~\ref{sec:scale-connected}, the marginal deviation from its
    (component) mean, depends on the graph, and will in general be
    different for each connected component. In this simple case, the
    (conditional) marginal variance is constant within each connected
    component, and equal $0.22/\kappa$, $0.25/\kappa$ and $\infty$,
    respectively.
\end{itemize}
We can resolve both these issues by scaling~\eref{eq4} similar to what
we did in \Sec{sec:scale-connected}, with two minor modifications.
\begin{enumerate}
\item We scale each connected component of size larger than one,
    independently as described in Section~\ref{sec:scale-connected}.
\item For connected components of size one, we replace it with a
    standard Gaussian with precision $\kappa$.
\end{enumerate}
This scaling gives a well defined interpretation of $\kappa$ and the
same typical (conditional) marginal variance within each connected
component, no matter the size. In our example, we obtain the following
scaled precision matrix:
\resizebox{.8\textwidth}{!}{
    $\displaystyle \textrm{Q}_{\text{scaled}} = \kappa \left( c_1
      \begin{bmatrix}{}
          2 & -1 & -1 &   &   &   \\
          -1 & 2 & -1 &   &   &   \\
          -1 & -1 & 2 &   &   &   \\
          &   &   & 0 & 0 &   \\
          &   &   & 0 & 0 &   \\
          &   &   &   &   &  0 \\
      \end{bmatrix} +c_2
      \begin{bmatrix}{}
          0 & 0 & 0 &   &   &   \\
          0 & 0 & 0 &   &   &   \\
          0 & 0 & 0 &   &   &   \\
          &   &   & 1 & -1 &   \\
          &   &   & -1 & 1 &   \\
          &   &   &   &   &  0 \\
      \end{bmatrix} +
      \begin{bmatrix}{}
          0 & 0 & 0 &   &   &   \\
          0 & 0 & 0 &   &   &   \\
          0 & 0 & 0 &   &   &   \\
          &   &   & 0 & 0 &   \\
          &   &   & 0 & 0 &   \\
          &   &   &   &   &  1 \\
      \end{bmatrix}
    \right) $}

where $c_1$ and $c_2$ are the scaling values for the two connected
components of size larger than one. Note, a normal prior with precision $\kappa$ is assigned to the singleton in order for its marginal variance to be the same as for the nodes in the connected components. Our recommandation is clear and
unambiguous. Therefore, due to the scaling, the precision
parameter has the same interpretation for all sub-graphs, also for the
singleton.

\begin{recommendation}
    We recommend to scale intrinsic CAR models defined with regard to
    disconnected graphs.
\end{recommendation}

\section{Linear constraints and normalising constants}
\label{sec:linear-constraints}

When using intrinsic models we always have to be careful not to introduce unwanted confounding.
Let $\mm{x}$ be the intrinsic CAR defined on
a connected graph, then linear predictors $\mm{\eta} = \mm{x}$ and
\begin{displaymath}
    \mm{\eta} = \mu\mm{1} + \mm{x}|(\mm{1}^T\mm{x}=0)
\end{displaymath}
are the same. In the first case, there is no intercept as it is
implicitly in the null-space of the precision matrix for $\mm{x}$. In
the second case, we explicitly define the intercept and remove it from
the intrinsic CAR model. We strongly prefer the second option, since
it makes the interpretation of each component explicit and reduce the
chance of misunderstanding and misspecifying the intrinsic CAR in more
complex scenarios than here.

When the graph is disconnected, we designate one intercept for each
connected component with size larger than one. Hence, we recommend to
use one sum-to-zero constraint for each connected component of size
larger than one. If one needs a connected component specific
intercept, we prefer to add it to the model explicitly rather than
implicitly.
\begin{recommendation} We recommend to use a sum-to-zero constraint
    for each connected component of size larger than one.
\end{recommendation}
 
Rue and Held~\cite[Ch~3]{book80} provide a strong case,  to interpret the normalizing 
constant for the proper part of the model and the improper part  as a diffuse Gaussian.
Assume the graph has $n_c$ connected components, each of
size $n_i$. Then the normalizing constant for the scaled intrinsic CAR
model~\eref{eq2}, will be
\begin{displaymath}
    Z_n(\kappa) = |\mm{R}|_*^{1/2}\prod_{i=1}^{n_c} Z_{n_i}(\kappa)
\end{displaymath}
where
\begin{displaymath}
    Z_{m}(\kappa) = 
    \begin{cases}
        (\kappa/(2\pi))^{1/2} & \text{if $m=1$} \\
        (\kappa/(2\pi))^{(m-1)/2}& \text{otherwise.}
    \end{cases}
\end{displaymath}
and $|\cdot|_*$ is a generalised determinant defined as the product of
all non-zero eigenvalues. In most cases, we only need the part of
$Z_n(\kappa)$ that depends on $\kappa$, hence we do not need to compute the generalised determinant nor carry the $2 \pi$ around.

\section{Application}
\label{sec:examples}

In this section we provide two applications, in the first one we show
how the scaling works with a disconnected graph, on lip cancer data
from Scotland. In the second application, we want to give a broad
picture of how scaling is related to the interpretation of the prior
assigned to the precision parameter, using lung cancer mortality data
for Tuscany Region (Italy).

\subsection{Scottish Lip Cancer data: a graph with three singletons}
\begin{figure}[tbp]
    \centering \subfloat[]{\includegraphics[scale=1.2,
        width=0.4\textwidth]{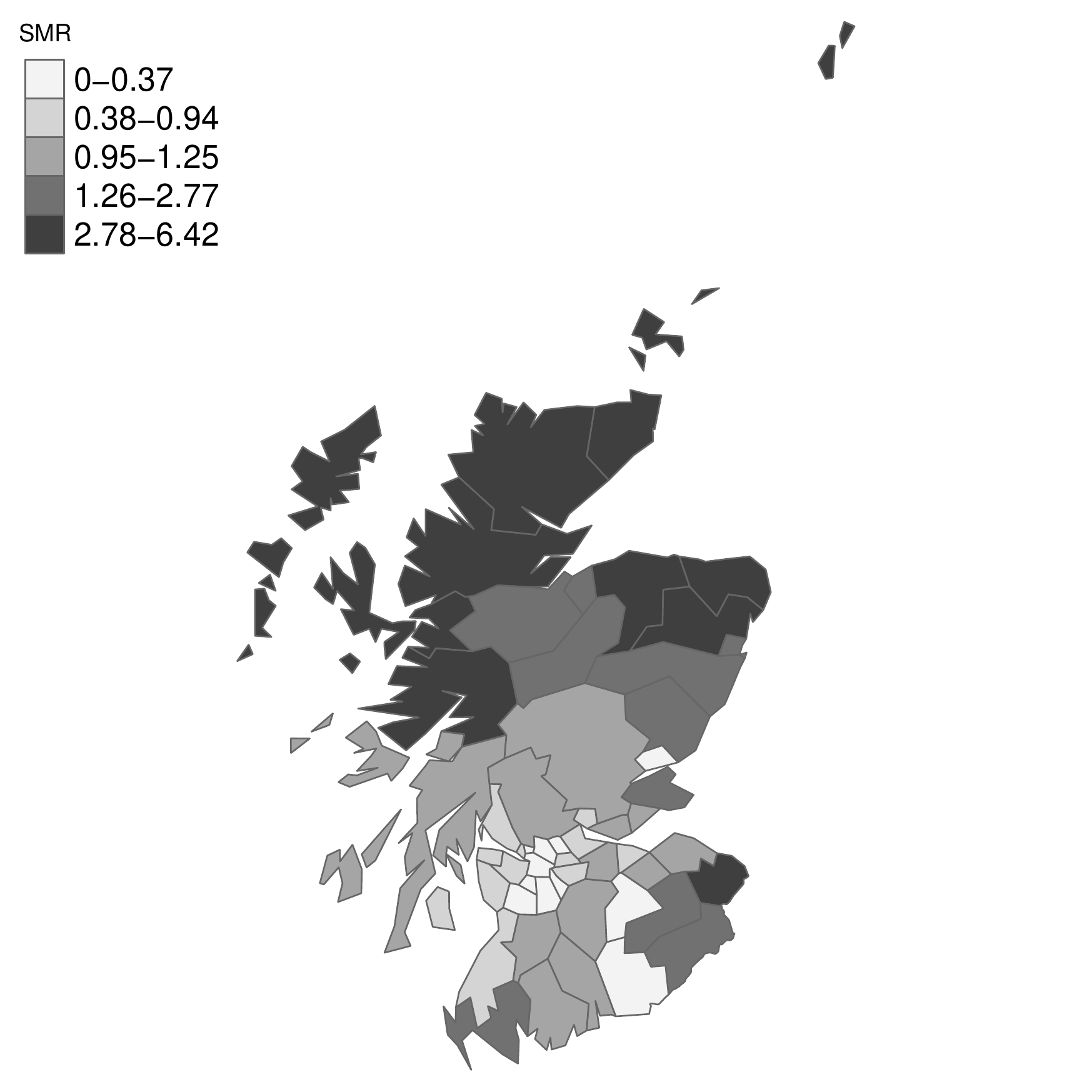}\label{fig:3}}
    \hfill
    \subfloat[]{\includegraphics[width=0.4\textwidth]{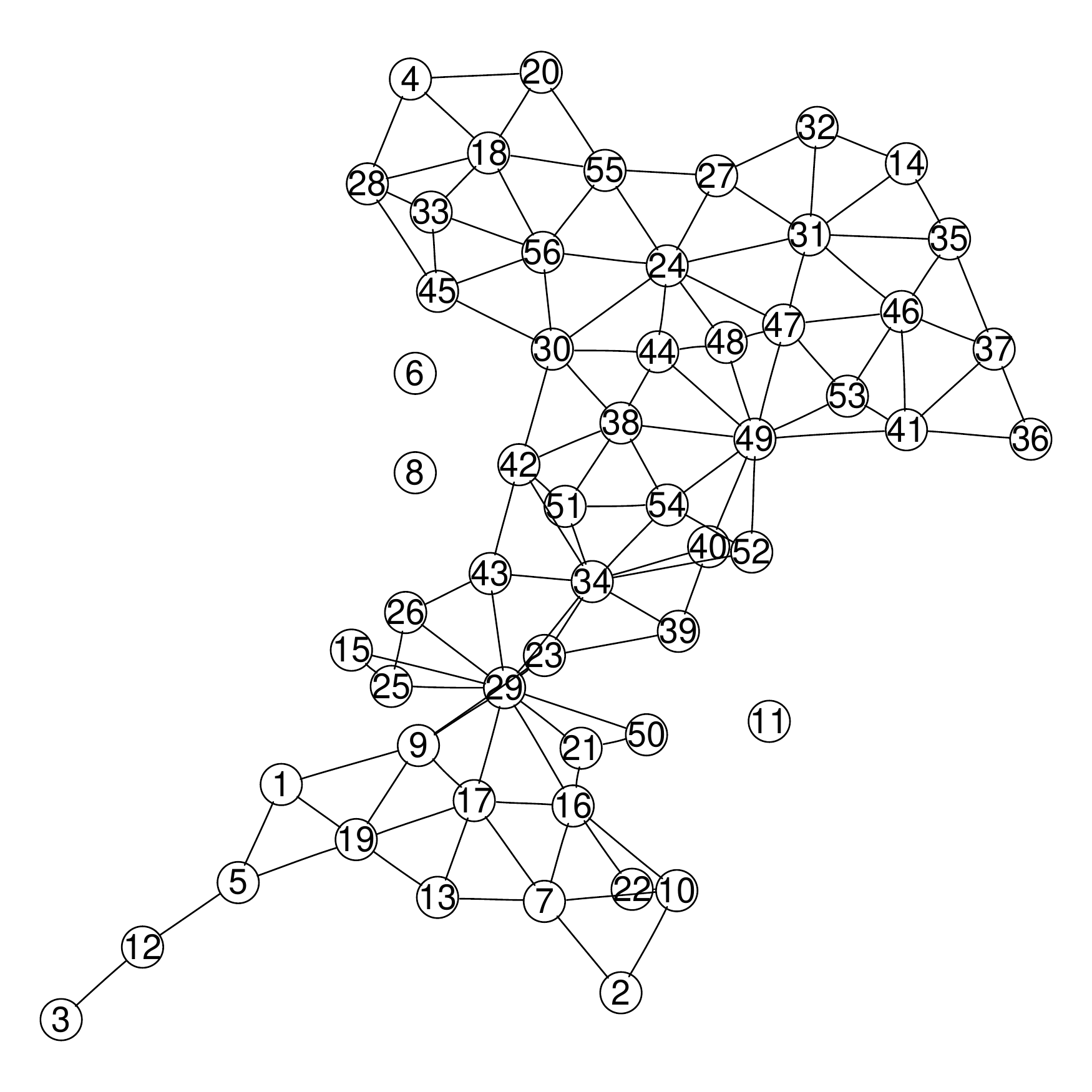}\label{fig:4}}
    \caption{Scotland map (a) and (b) the disconnected graph induced
        by the map.}
\end{figure}

The data are counts of lip cancer cases registered in 56 Scottish
counties during years 1975-1980. We want to smooth the observed
Standard mortality ratios (SMR), Figure~\ref{fig:3}.
We generate a graph by assuming the counties are the nodes, with edges
connecting counties sharing borders, Figure~\ref{fig:4}. Three
counties are islands (Orkneys, Shetland and the Outer Hebrides),
therefore we are left with three singletons in our graph. Breslow et
al.~\cite{breslow1993} analyzed the spatial dependency using an
intrinsic CAR model defined on a connected graph, obtained by editing
new edges to connect the islands.
The popular WinBUGS software treats singletons as non-stochastic nodes
and sets the associated random effects to zero by default (see
\texttt{GeoBUGS} manual\cite{geobugs} page 18). From the perspective
of developing user friendly software to fit Bayesian models via Markov
Chain Monte Carlo (MCMC) algorithms this seems a safe strategy: if
singletons are taken as stochastic nodes, with consequent improper
$\pi(x_i) \propto const$, this may lead to poor mixing and extremely
slow convergence especially when a zero count is observed; see the
discussion in Section~\ref{sec:scale-disconnected}.

We argue that removing the singletons is needless for the definition
of a suitable intrinsic CAR model. Our recommended solution is to avoid this and to assign the island-specific random effects a normal prior with zero mean and variance equal to $\kappa^{-1}$
\cite{Wakefield2007}.

Assuming vectors $\mm y$ and $\mm E$ are, respectively, observed and
expected lip cancer cases during the study period, covariate $\mm z$
is the ``percentage of the population engaged in agriculture, fishing,
or forestry''(AFF) and $\mm{r}$ the unknown relative risks, the model
is:
\begin{eqnarray}
  \label{eq:model-lip-cancer1}
  \mm{y} &\sim & \text{Po}\left(\mm{E} \mm{r}\right) \\
  \log(\mm{r}) & =&  \alpha + \beta \mm{z} + \mm{x}  \\
  \pi(\mm{x} \mid \kappa)& \propto & \frac{1}{Z_n(\kappa)}
                                     \exp\left(
                                     -\frac{\kappa}{2} \mm{x}^{T} \mm{R} \mm{x}
                                     \right).
                                     \label{eq:model-lip-cancer2}
\end{eqnarray}
In a direct application of the CAR model, the (\emph{unscaled}) structure matrix $\mm{R}$ in ~\eref{eq:model-lip-cancer2} contains the number of neighbours, $n_i$, in position $(i,i)$ and values $-1$ in positions $(i,j)$, $i \sim j$. If $i$ is a singleton, then $\mm{R}[i,i]=0$ and the prior for $x_i$ is constant. According to the recommendations in Section~\ref{sec:scale-disconnected}, we use the {\emph{scaled} version of ~\eref{eq:model-lip-cancer2}, meaning that we scale the connected component of the graph of size larger than one (mainland) and assign a $\mathcal{N}(0,\kappa^{-1})$ prior for
each of the three singletons (islands). The scaling is coded in the \texttt{inla.scale.model()} function; in practice it is sufficient to flag as true the \texttt{scale.model} option when specifying the latent model \texttt{$f()$} in the package \texttt{R-INLA} \cite{art451,art522,art632} (see the code in the supplementary material~\ref{sec:sm})

The benefit of scaling is well illustrated by comparison
against the unscaled version of the intrinsic CAR model. For the sake of comparison, for both scaled and unscaled models we assume a gamma with shape \texttt{1} and rate \texttt{5e-5} for the precision $\kappa$ and apply a sum-to-zero constraint to the connected component of the graph (according to our recommendation given in Section~\ref{sec:linear-constraints}).

\begin{figure}[tbp]
    \begin{tabular}{ccc}
      \includegraphics[width=0.3\linewidth]{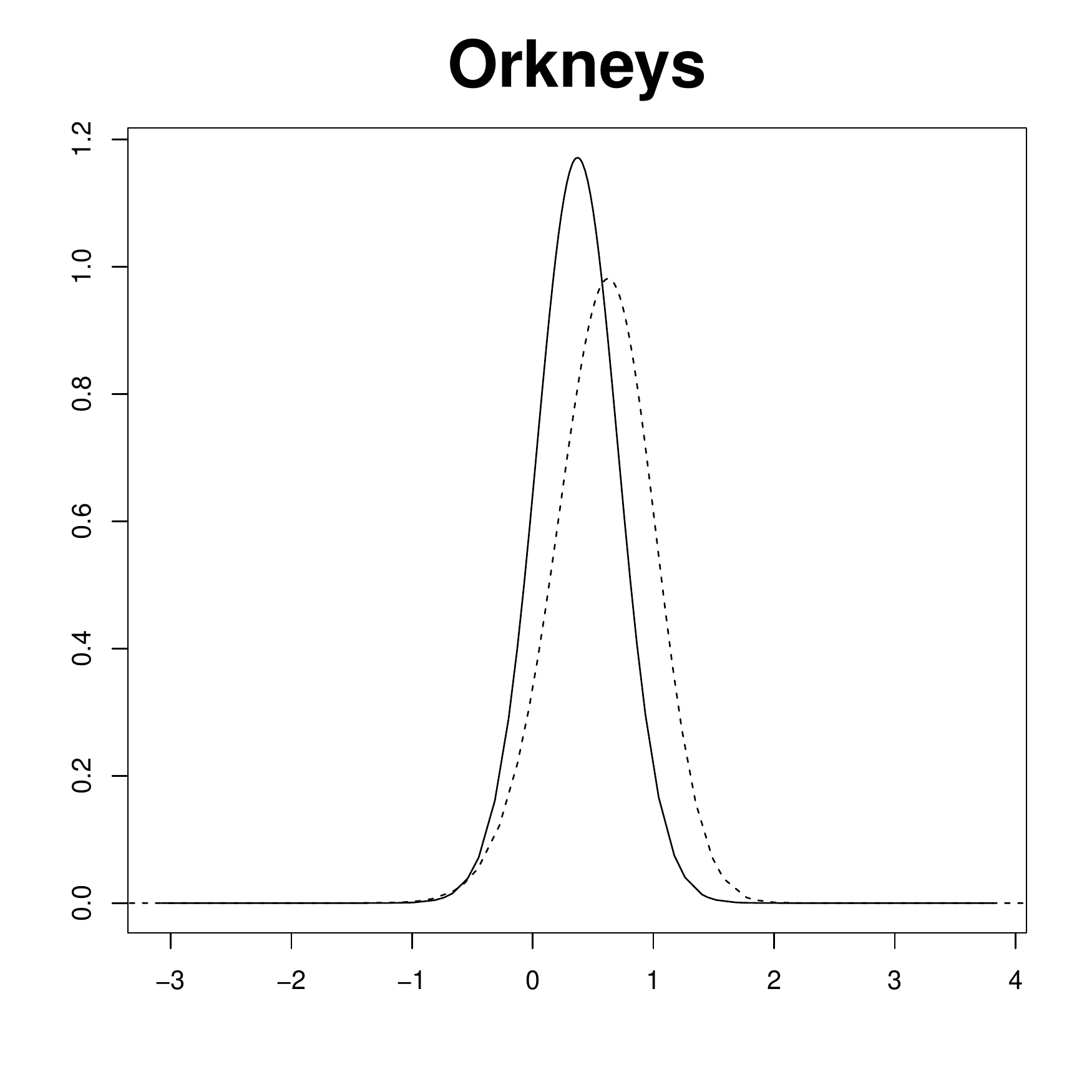}
      & \includegraphics[width=0.3\linewidth]{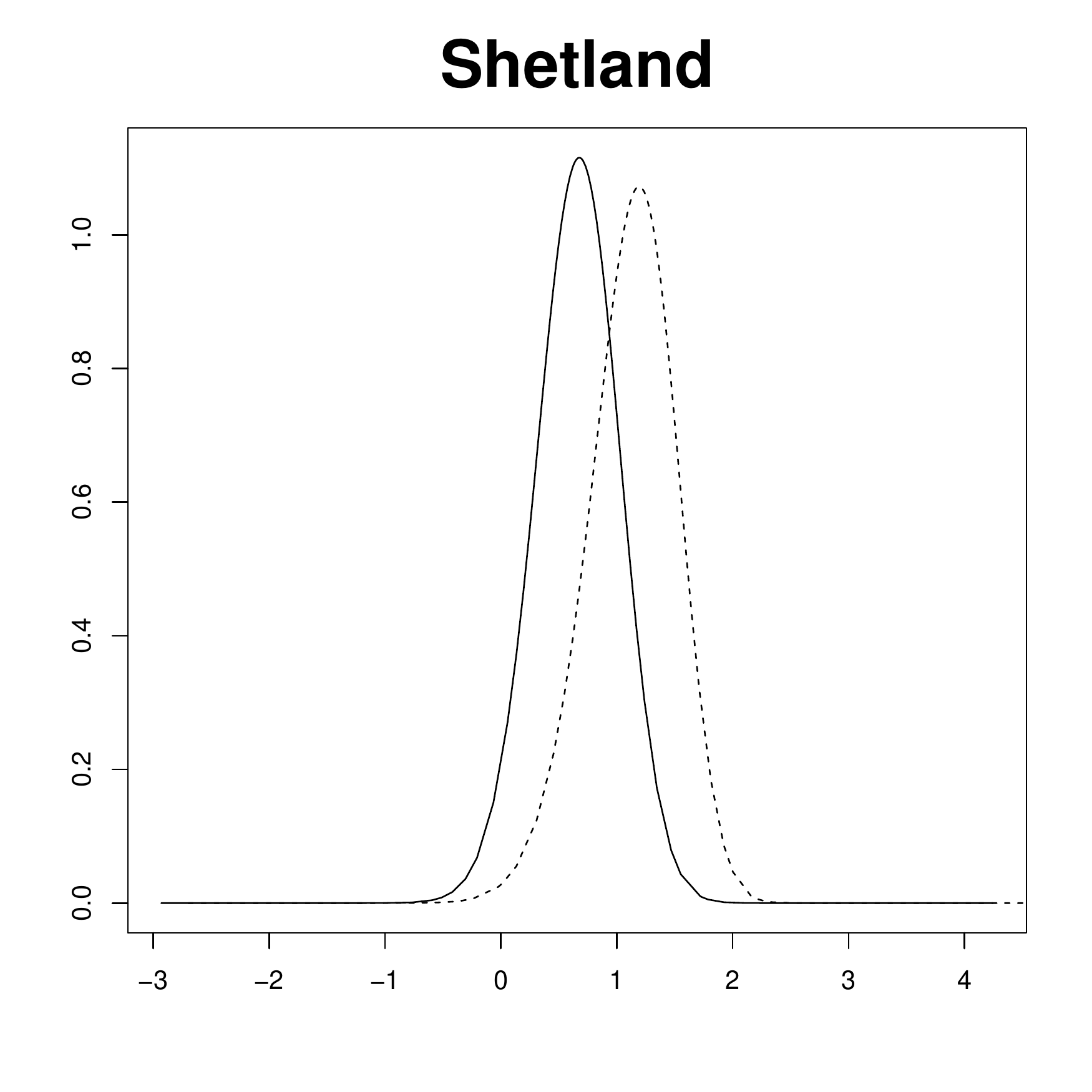}
      & \includegraphics[width=0.3\linewidth]{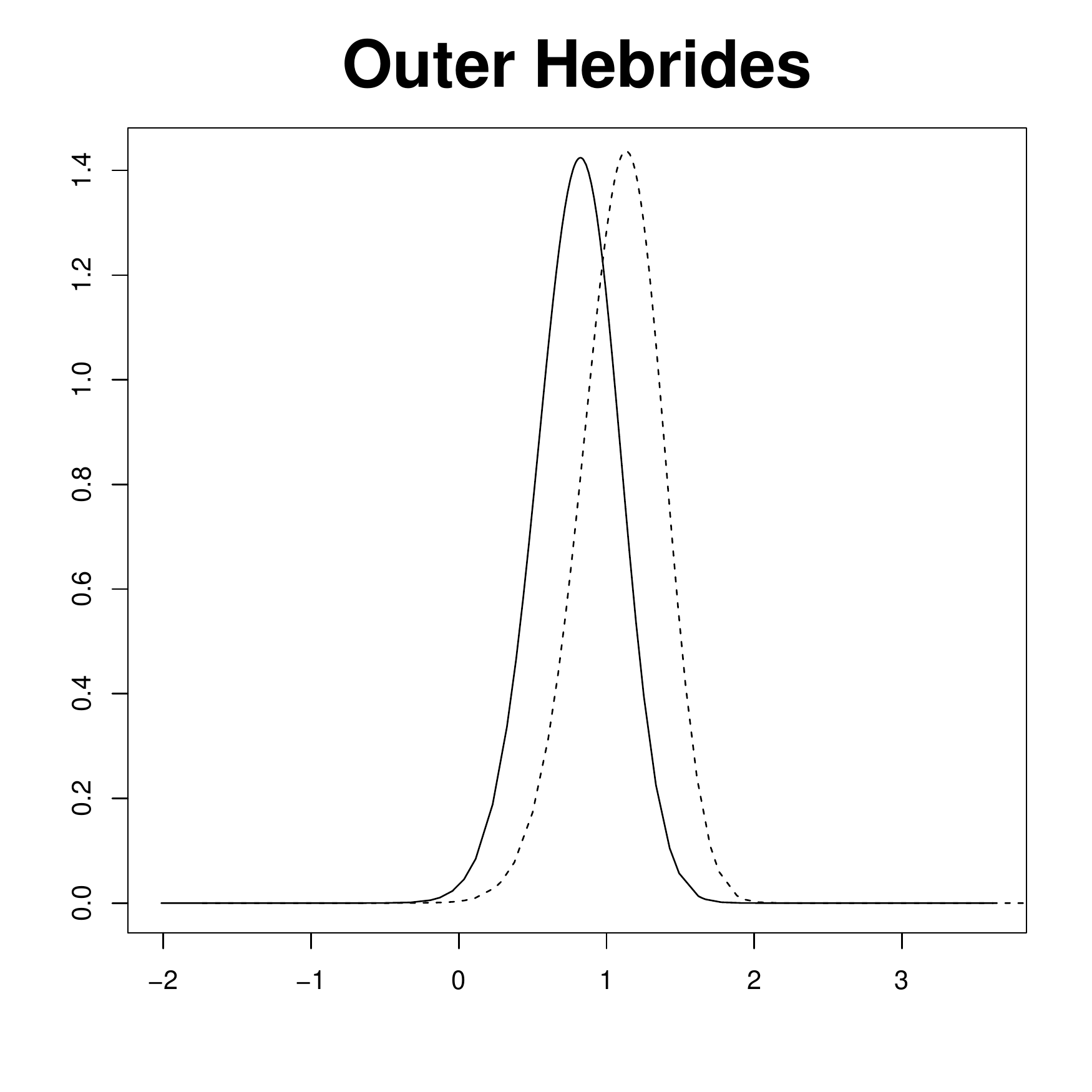}\\
      \includegraphics[width=0.3\linewidth]{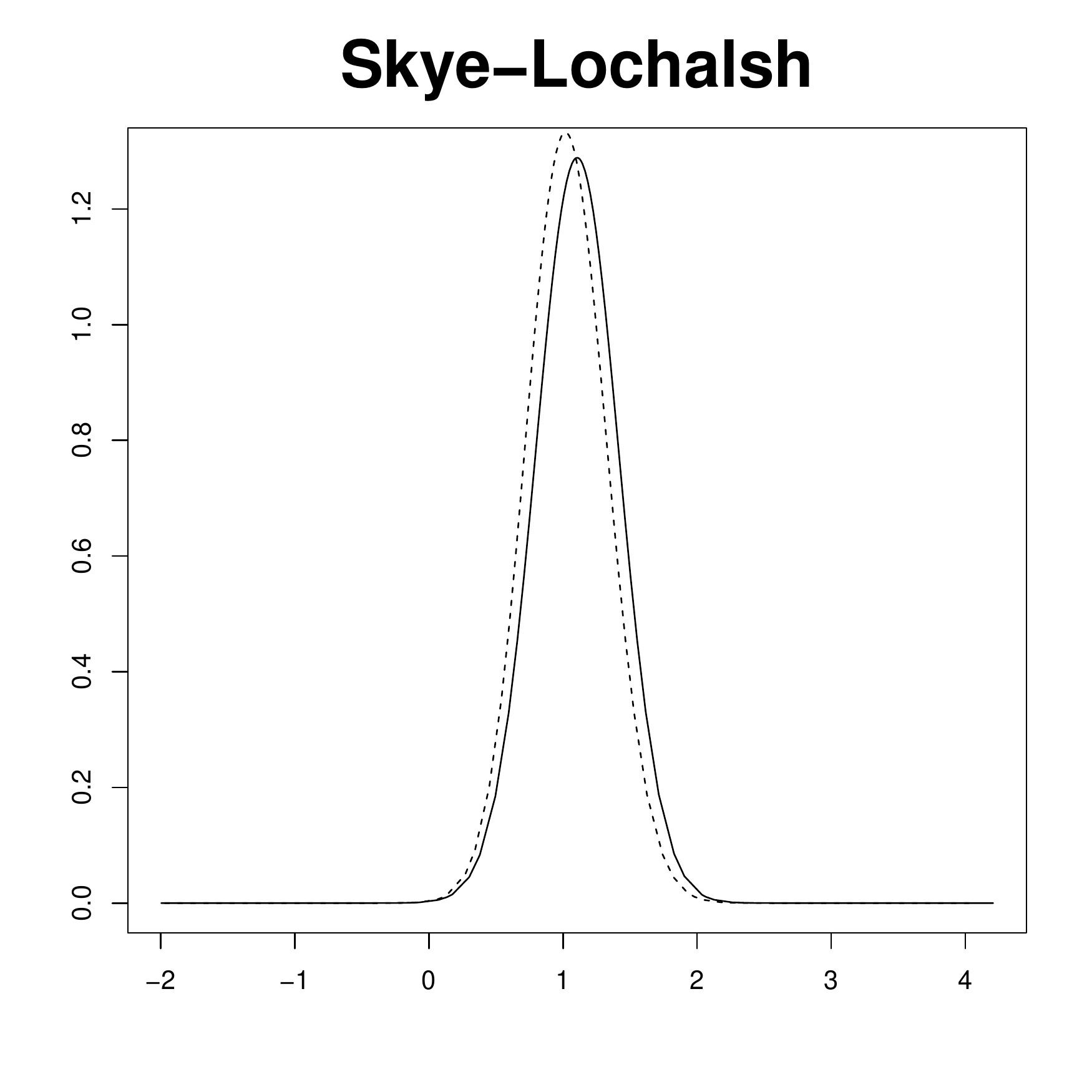}
      & \includegraphics[width=0.3\linewidth]{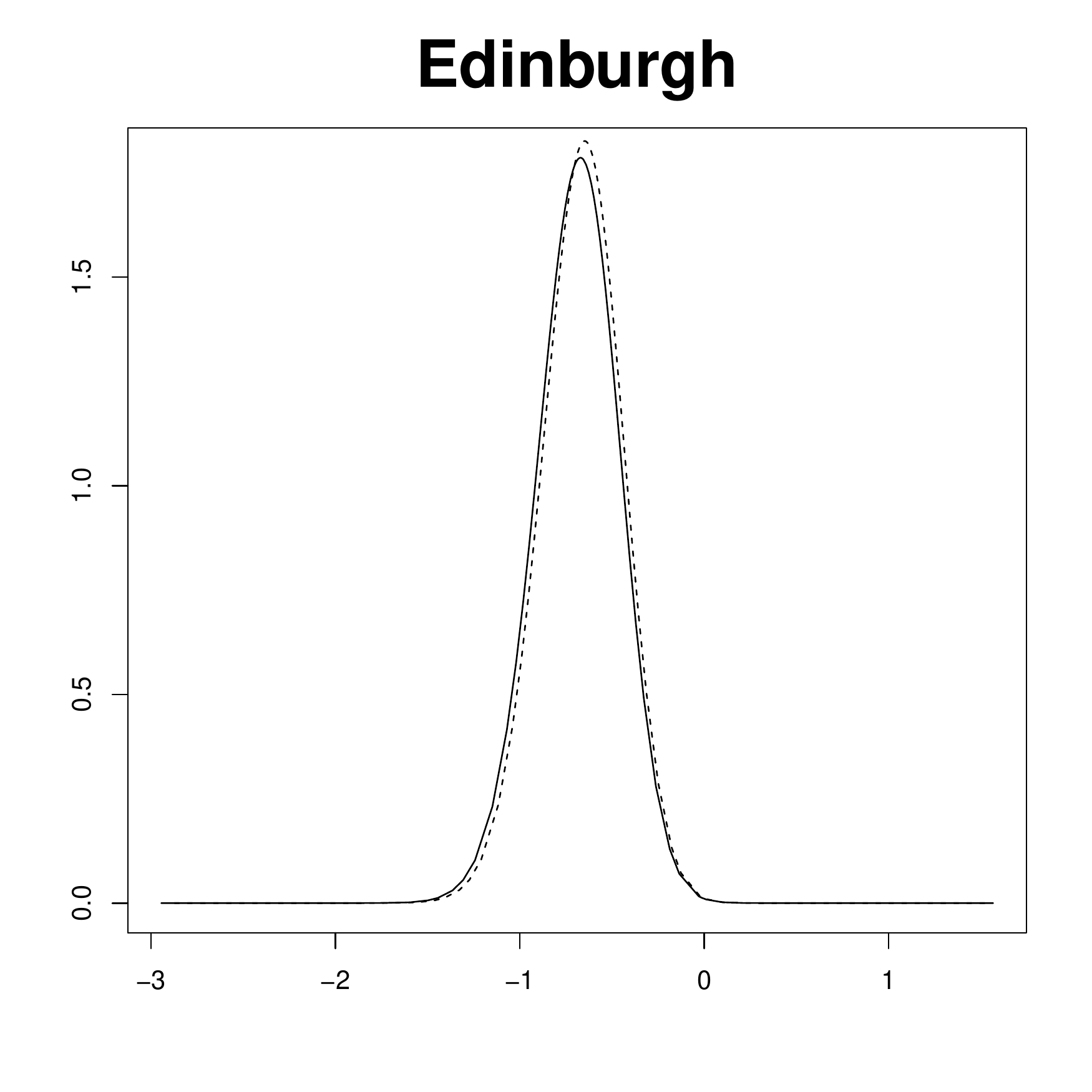}
      &  \includegraphics[width=0.3\linewidth]{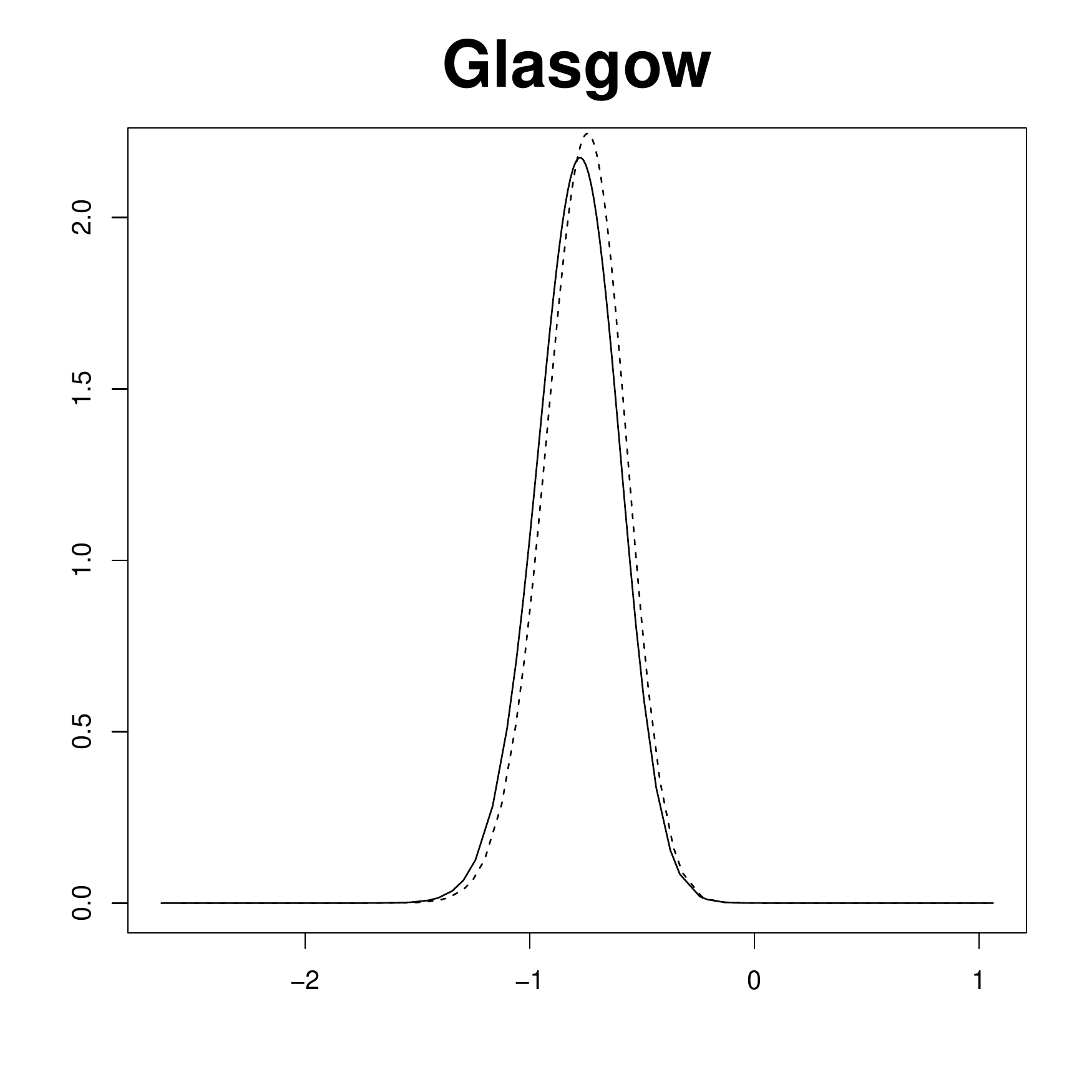} \\
    \end{tabular}
    \caption{The effect of scaling the disconnected graph. Upper
        panels show the marginal posterior (in the linear predictor
        scale) for the three singletons random effects, $x_6$
        (Orkneys), $x_8$ (Shetland) and $x_{11}$ (Outer Hebrides);
        lower panels show the marginal posterior for three nodes in
        the connected component of the graph, $x_1$ (Skye-Lochalsh),
        $x_{45}$ (Edinburgh) and $x_{49}$ (Glasgow). The marginals
        from the unscaled model (dashed lines) are less shrunk towards
        $x_i=0$ than the marginals from the scaled model (solid
        line).}
    \label{fig:5}
\end{figure}

\Fig{fig:5} displays the marginal posterior for six random effects
fitted under the scaled (solid line) and unscaled (dashed line) model,
using \texttt{R-INLA}. In the scaled model, the
hyper-parameter $\kappa$ has a clear interpretation as a typical precision. In other words, the (conditional) marginal variance within each of the four components of the Scotland graph (the mainland and the three singletons) is proportional to $\kappa^{-1}$. In the unscaled model, the (conditional) marginal variances are different in each node of the graph. This issue is reflected in the large deviations in the top panels of \Fig{fig:5}, referring to the three singletons (Orkneys, Shetland and
Outer Hebrides). Note that the island-specific random effects, estimated by the unscaled model, are less shrunk towards no effect, $x_i=0$, than those estimated by the scaled model. On the other hand, the posterior for the three random effects belonging to the connected component of the graph (bottom panels of \Fig{fig:5}) are essentially unchanged between the two models. Results for the other nodes in the connected component are similar and not shown here.
The different shrinkage properties of the scaled and unscaled models are confirmed by looking at posterior summaries for the main model parameters in Table \ref{tab:1}. Though the fixed effects $\alpha$ and $\beta$ are almost unchanged, the relative risks for the singletons are more extreme under the unscaled model than the scaled one.

\begin{table}[tbp]
    \small
    \centering
    \begin{tabular}{llllll}
      \hline
      Parameter & Mean & Standard & 2.5\% & Median & 97.5\%\\
                &        &deviation & &  & \\
      \hline
      \\
      \multicolumn{5}{l}{\emph{Scaled model }} & \\ \\
      $\kappa$ (Marginal variance)& 3.97 & 1.17 & 2.16 & 3.81 & 6.69  \\
      $\alpha$ (Intercept) & -0.25 & 0.13 & -0.50 & -0.25 & -0.00 \\ 
      $\beta$ (AFF)  & 0.37 & 0.13 & 0.09 & 0.37 & 0.62  \\
      $r_6$ (Orkneys) & 2.87 & 0.90 & 1.42 & 2.77 & 4.93 \\
      $r_8$ (Shetland) & 2.06 & 0.73 & 0.95 & 1.95 & 3.80 \\ 
      $r_1$ (Outer Hebrides) & 2.32 & 0.63 & 1.28 & 2.26 & 3.75 \\ 
      \hline 
      \\
      \multicolumn{5}{l}{\emph{Unscaled model }} & \\ \\
      $\kappa$ (Marginal variance) & 2.26 & 0.70 & 1.18 & 2.15 & 3.91 \\ 
      $\alpha$ (Intercept) & -0.26 & 0.12 & -0.50 & -0.27 & -0.02 \\ 
      $\beta$ (AFF) & 0.36 & 0.13 & 0.09 & 0.37 & 0.62 \\
      $r_6$ (Orkneys) & 3.54 & 1.20 & 1.58 & 3.40 & 6.27\\ 
      $r_8$ (Shetland) & 3.26 & 1.18 & 1.36 & 3.11 & 5.96  \\  
      $r_1$ (Outer Hebrides)& 3.07 & 0.83 & 1.66 & 2.99 & 4.89\\  
      \hline 
    \end{tabular}
    \caption{Posterior summaries for scaled (top) and unscaled
        (bottom)
        versions of model~\eref{eq:model-lip-cancer1}-\eref{eq:model-lip-cancer2}.}
    \label{tab:1}
\end{table}

\subsection{Tuscany Lung cancer mortality: a graph split in
    sub-graphs}
\label{sec:exp2}
In Section~\ref{sec:intro}, we assert that the scaling plays an
important role in the interpretation of the assigned prior on
hyper-parameter $\kappa$, because it allows to have priors with the
same interpretation on different underlying graphs.

In this second application, we want show the combination of the
scaling and precision hyper-prameter priors, on an underlying
disconnected graph. To show what we intend, we will introduce briefly
a new parametrization for an intrinsic conditional autoregressive model
and the concept of penalized complexity riors, see Simpson et al.~\cite{art631}

Riebler et al.~\cite{art585} defined an alternative Besag-York-Mollie
(BYM) \cite{art149} parametrization, for the intrinsic conditional
autoregressive, named \emph{bym2} (within \texttt{R-INLA}). The
\emph{bym2} parametrization specifically accommodates scaling for
connected or disconnected graph. In \emph{bym2}, the random effect is
$x=v+u_*$, where $v$ is the spatially unstructured component and $u_*$
is the scaled spatially structured component (i.e. the scaled intrinsic CAR model).

The random effect is re-parametrized as
\begin{equation}
\label{eq:by2}
    \mm{x}=\frac{1}{\sqrt{\tau_x}}\left(\sqrt{1-\phi}\mm{v}+\sqrt{\phi}\mm{u}_*\right)
\end{equation}
with a covariance matrix
\[
    \text{Var}(\mm{x}|\tau_x)=\tau_x^{-1}\left((1-\phi)\mm{I}+\phi\mm{
          Q_{*}^-}\right)
\]
The total variance is expressed by a mixing parameter $\phi$
($0 \leq \phi \leq 1$) that measures the proportion of marginal
variance due to the structural spatial effect. In addition, $\tau_x$
represents the precision of the marginal deviation from a constant
level, without regard for any type of underlying graph. Finally,
$\mm{Q_{*}^-}$ indicates the generalised inverse of the precision matrix.
If $\phi=0$ the model is based solely on overdispersion, while if
$\phi=1$ the model coincide with a Besag model: a pure structured
spatial effect. Now, the specification of priors for $\phi$ and
$\tau_x$ follows a penalized complexity priors approach (pc-priors).
The pc-priors framework follows four principles: 1. \emph{Occam's
    razor}- simpler models should be preferred until there is evidence
from more complex. 2. \emph{Measure of complexity}- the
Kullback-Leibler distance is used to measure increased complexity. 3.
\emph{Constant rate of penalization}- the deviation from simpler model
has a constant decay rate. 4. \emph{User defined scaling}- the user
has a clear idea of a sensible size for parameters or on their
transformations. Therefore, pc-priors are defined as informative
priors that will penalize departure from a base-model. In this setting
a base model presents a constant relative risk surface, therefore no
spatial variation, opposed to a \emph{complex} model, that shows
spatial variation. The clarity gained by the pc-priors framework,
allows the user to comprehensively state priors in terms of beliefs on
$\phi$ and $\tau_x$~\cite{art585}, where $\phi$ and $\tau_x$ are
declared by choices of $U$ and $\alpha$ in $Pr(\phi <U)=\alpha$ and
$Pr(1/\sqrt{\tau_x)} <U)=\alpha$, respectively. The two probabilities
provide the user with an easy way to define an upper bound to what the user 
thinks as \emph{tail-event}, and to assign an $\alpha$ to this event
it. The advantage of the pc-priors is the invariance to
parametrization, that's why they are very hand in situation with
disconnected scaled graphs. We will show how these priors are declared
and the results obtained, by using lung cancer deaths in women dataset
(2585 total cases), collected for each municipality between 1981-1989,
for the Tuscany Cancer Atlas. Tuscany presents two small islands, that
have separate municipality and we will consider them as singletons -
Capraia and Giglio Isles - and Elba Isle composed of 8 municipalities.
In~\Fig{fig:6} we plotted the standardised mortality ratio on Tuscany
map, and the induced graphs composed of two major connected components
and two singletons,~\Fig{fig:7}. Note that SMRs are higher for
northern municipalities and for Elba Island , where iron mine and
steel mills were active since 1905 \cite{artbig}. The SMR variance
range is 0-100.

We fitted a Poisson regression without covariates and two pc-priors
choices: the default values embedded in
\texttt{R-INLA}(i), and then with some more informative priors
values(ii).  
In the default setting $\phi$ has $U=0.5$ and
$\alpha=0.5$, which assumes that the unstructured and structured
random effects account equally to the total variability and
$\tau_x$ has $U=1$, a prior that corresponds to a marginal standard deviation for $\mm{x}$ of $0.31$ and therefore to a residual relative risk smaller than 2, see Simpson et al.~\cite{art631}. 
For our second choice, we specify $Pr(1/\sqrt{\tau_x}>0.1/0.31)=0.05$, hence we assign a 95\% to have a marginal standard deviation of $0.1$.
while maintaining $\phi$'s prior as default.
For $\tau_x$ the choice of $U$ and $\alpha$ is less intuitive than the
mixing parameter, but $\tau_x$ is the marginal precision related to
the residual relative risk. This means that we assign a 95\% of 
having a residual relative risk smaller than $1.17$. 

In table~\ref{tab:2}, the intercept estimates stay unchanged, we see a
slight change for the precision, while $\phi$ posterior marginal is
supporting a major influence on non-spatial variability with a narrow
credible interval, in both instances. Based on the Deviance
Information Criterion (DIC) the first choice prior model is the one
to prefer. However, as this can be seen as a sensitivity analysis, by
twisting the precision prior we observed similar results. We are less
concerned in this example with islands random effects, because we know
that due to proper scaling they are shrank toward zero random effects.

\begin{figure}[tbp]
    \subfloat[]{\includegraphics[width=0.4\textwidth]{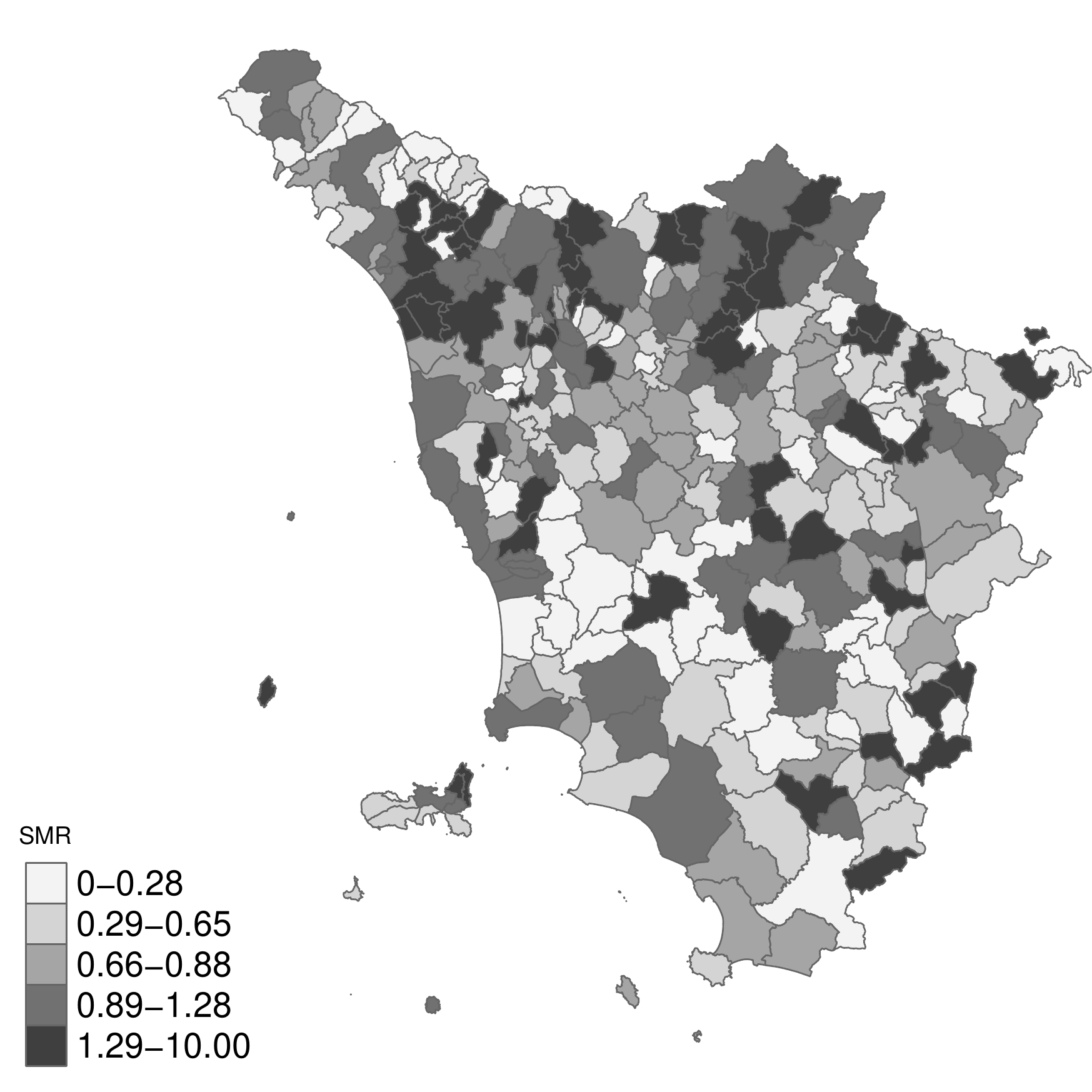}\label{fig:6}}
    \hfill
    \subfloat[]{\includegraphics[width=0.4\textwidth]{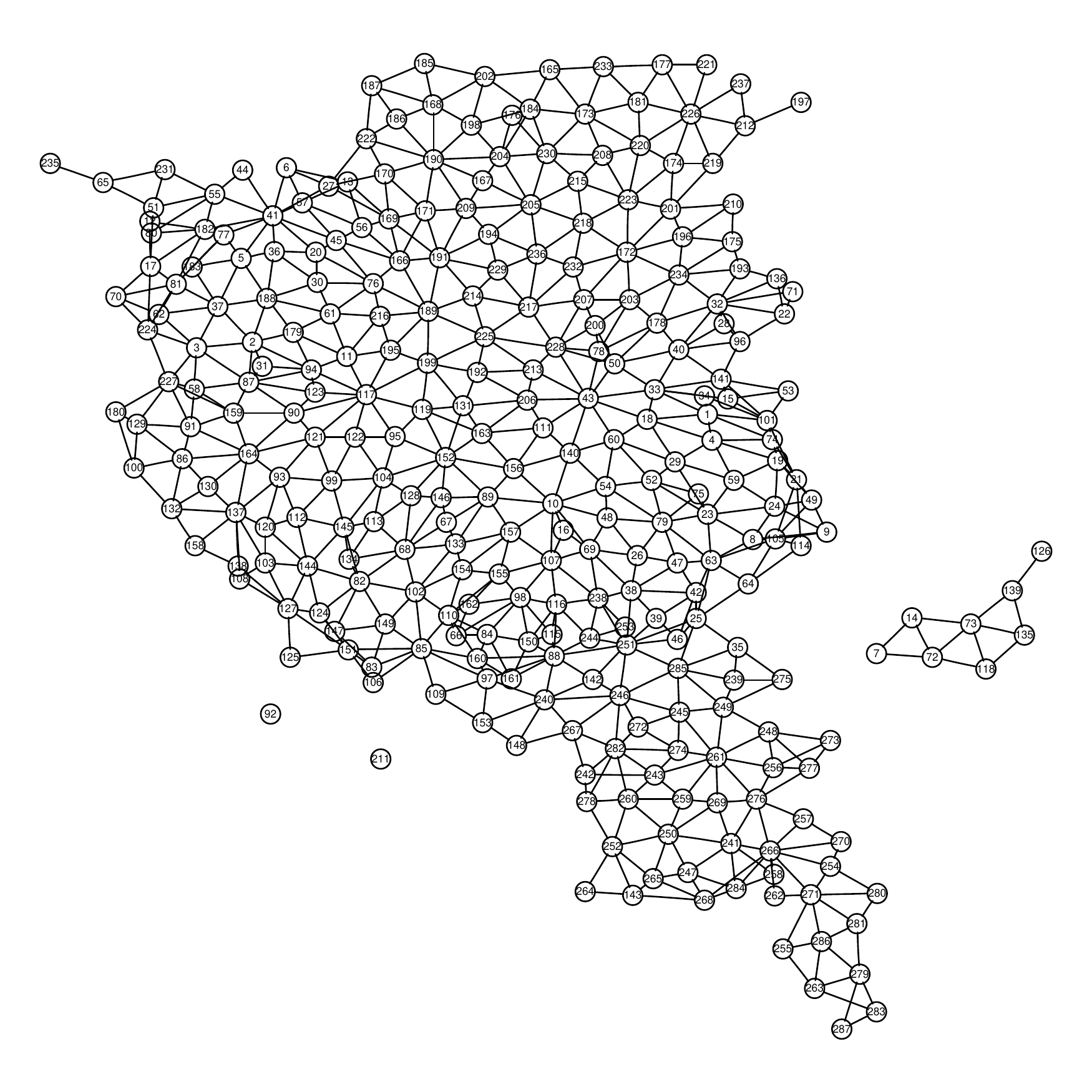}\label{fig:7}}
    \caption{Tuscany map (a) and the disconnected graph induced by
        the map (b).}
    \label{fig:6/7}
\end{figure}

\begin{table}[htbp]
    \small
    \centering
    \begin{tabular}{llllll}
      \hline
      Parameter & Mean & Standard & 2.5\% & Median & 97.5\%\\
                &        &deviation & &  & \\
      \hline
      \\
      \multicolumn{5}{l}{\emph{Prior for $\tau_x, U=1$, $\alpha=0.01$}} & \\ \\
      $\alpha$(Intercept) & 1.12 & 0.06 & 0.99 & 1.12 & 1.24 \\ 
      $\tau_x$(Precision) & 1.31 & 0.22 & 0.93 & 1.30& 1.79 \\ 
      $\phi$(Mixing parameter) &0.24 & 0.12&0.06 &0.22&0.54 \\
      DIC & 1114.21   & &&&\\
      \hline 
      \\
      \multicolumn{5}{l}{\emph{Prior for $\tau_x, U=0.1/0.31$, $\alpha=0.05$}} & \\ \\
      $\alpha$(Intercept) & 1.126 & 0.06 & 1.00 & 1.124 & 1.128\\
      $\tau_x$(Precision) & 1.38& 0.22 & 0.99 & 1.37 & 1.87 \\ 
      $\phi$(Mixing parameter)&0.26 &0.13 &0.06 &0.22 &0.52 \\
      DIC &1117.38    &&&&\\
       
      \hline 
    \end{tabular}
    \caption{
        Posterior summaries for intercept and hyperparameters
        and Deviance Information Criterion (DIC), under the
        two pc-priors choices
        for  marginal variance $\tau_x$ on Tuscany lung cancer data,
        while holding default mixing parameter priors.}
    \label{tab:2}
\end{table}

\section{Summary}
\label{sec:summary}

We motivated the definition of intrinsic CAR models for disconnected
graphs under two main recommendations: scaling the precision structure
and applying sum-to-zero constraints on the connected components of
the graph. Scaling the precision structure sets the typical marginal
variance to $\kappa^{-1}$) in each component of the graph, where $\kappa$ is the precision of the intrinsic CAR model. This immediately suggests a fair prior for random effects associated to the singletons in a disconnected graph in terms of a normal with zero mean and variance $\kappa^{-1}$. The advantage is that the prior assigned to $\kappa$ has the same interpretation regardless of the particular structure of the graph. 

We applied this strategy to a disease mapping example on lip cancer in
Scotland, using the natural disconnected graph with three island
regions. In this example we emphasized overfitting of the unscaled
model compared to the scaled one. In general, the extent to which the
unscaled intinsic CAR leads to overfitting should depend on the
structure of the graph, the sample size and the prior assigned to
$\kappa$. When data contain little information about the disease risk
for people living in the area (e.g. small regions with low expected
counts, which is not the case with the lip cancer data), the prior
$\pi(\kappa)$ may have large impact on the analysis. In this
situations larger deviation between scaled and unscaled intrinsic CAR
models must be expected, as the unscaled one has no control on the
marginal variance, hence no control on the impact of $\pi(\kappa)$,
whereas the scaled one provides good intuition of $\kappa$.

In the second example we went further and combined the scaled strategy
with an alternative parametrization and the application of informative
priors (penalized complexity priors). We recommend always to use a scaled version
for fitting disease mapping model for connected and disconnected
graphs and we demonstrated that the modified BYM model (\emph{bym2})
has clear and unambiguous parameter interpretation that are invariant
to the underlying graph.

\newpage

    \section*{Supplementary Material:\\
INLA code to implement  BYM2 models for
        disconnected graph}
    \label{sec:sm}
    We show how to implement the model BYM2 for using R
    -package INLA, using data from Scotland lip cancer.
\begin{lstlisting}
 #Load the R-package
library(INLA) 

# load data
data(Scotland)

# show the first lines
head(Scotland)
#  Counts   E  X Region
#1      9 1.4 16      1
#2     39 8.7 16      2
#3     11 3.0 10      3
#4      9 2.5 24      4

# read the graph structure
graph.scot = system.file("demodata/scotland.graph", package="INLA")
g = inla.read.graph(graph.scot)

# remove the edges to actually obtain the singletons, 
# these are nodes 6,8 and 11 (Orkneys, Shetland and 
# the Outer Hebrides)
id.singletons <- c(6,8,11)
G<-inla.graph2matrix(g)
G[id.singletons,]<-0
G[,id.singletons]<-0

#generates the disconnected graph 
g.disc <- inla.read.graph(G)


#sepcify the latent structure usig a fromula object 
formula.bym2 = Counts~ 1+f(Region, model='bym2',
                           scale.model=TRUE,
                             adjust.for.con.comp=TRUE,
                              graph=g.disc,
                               hyper=list(
                                 phi=list(
                                      prior='pc',
                                        param=c(0.5, 0.5)),
                                    prec=list(
                                      prior='pc.prec',
                                      param=c(0.2/0.31,0.05),
                                      initial=5)))
# call to the inla function
result= inla(formula.bym2, family="poisson", E=E, data=Scotland)
\end{lstlisting}

First we  install the INLA package in R, with the command:
\begin{lstlisting}
install.packages("INLA", 
repos="https://inla.r-inla-download.org/R/testing")
\end{lstlisting}
In this paper we used INLA version0.0-1493893899 and we show the \emph{bym2} using Scotland data and graph, that are 
embedded  in the package distribution.
In line 2--8, we load the library and the Scotland data, and then inspect the first rows. The dataset is composed by four variables (line 9) :
``Counts''- the number of lip cancer recorded; ``E'' the expected number of lip cancer;  ``X'' the percentage of the population engaged in agriculture fishing or forestry and 
``Region''  for the  county. 
Then we read the graph associated with Scotland, lines 16--17.  Because the graph is connected, we need to manually set the islands as singletons, lines 22-28. To do this we
change the graph in to a matrix object and then we assign 0 in the respective positions, and re-transform the matrix in a graph. 

In lines 32--43 we define the model structure in terms of an formula object. The \texttt{f()} function is used to specify the random effect in INLA. We specify the model in equation~\ref{eq:by2}, by passing the as first argument the "Region'', then declaring the model we are going to use, in this case \texttt{bym2}. 
We flag as true the options: \texttt{scale.model} to scale the graph and the \texttt{adjust.for.con.comp} to adjust for more than one connected component. We then provide the graph and in a list of arguments for the hyper parameters. In lines 37--39,  we declare $\phi$ as a pc-prior and the two values for $U$ and $\alpha$. Similarly, in lines 40--42 we do the same for the precision parameter and in argument \texttt{param} we set first $U$ and then $\alpha$. The last argument of the hyper parameter list is \texttt{initial} that sets  a value for the numerical optimisation start in INLA algorithm.

Finally we are ready to call function \texttt{inla}, with the formula,  data and the likelihood. The object result can be inspected by typing \texttt{summary(result)} and \texttt{ plot(result)}, while posterior summaries are stored in \texttt{results\$summary.fixed} and  \texttt{result\$summary.random}.

\section*{Acknowledgement}
    We thank Dr M.\ A.\ Vigotti (University of Pisa) for having made
    available the dataset from the Tuscany Atlas of Mortality
    1971-1994.
Massimo Ventrucci has been partially supported by the PRIN2015 (EphaStat) project founded 
by the Italian Ministry for Education, University and Research.

\bibliography{mybib}

\begin{thebibliography}{10}

\bibitem{art149}
J.~Besag, J.~York, and A.~Molli\'{e}.
\newblock {B}ayesian image restoration with two applications in spatial
  statistics (with discussion).
\newblock {\em Annals of the Institute of Statistical Mathematics},
  43(1):1--59, 1991.

\bibitem{artbig}
A.~Biggeri, D.~Catelana, and E.~Dreassia.
\newblock The epidemic of lung cancer in tuscany (italy): A joint analysis of
  male and female mortality by birth cohort.
\newblock {\em Spatial and Spatio-temporal Epidemiology}, 1(1)(31–40), 2009.

\bibitem{Hodges2003}
James~S. Hodges, Bradley~P. Carlin, and Qiao Fan.
\newblock On the precision of the conditionally autoregressive prior in spatial
  models.
\newblock {\em Biometrics}, 59(2):317--322, 2003.

\bibitem{Knorr}
L.~Knorr-Held.
\newblock Some remarks on gaussian markov random field models for disease
  mapping.
\newblock In P.~Green, N.~Hjort, and S.~Richardson, editors, {\em Highly
  structured stochastic systems}, pages 260--264. Oxford University Press,
  Oxford, 2002.

\bibitem{book116}
A.~B. Lawson.
\newblock {\em Bayesian Disease Mapping: Hierarchical Modeling in Spatial
  Epidemiology}.
\newblock Chapman \& Hall/CRC Interdisciplinary Statistics. Chapmann \&
  Hall/CRC, 2nd edition, 2013.

\bibitem{art522}
T.~G. Martins, D.~Simpson, F.~Lindgren, and H.~Rue.
\newblock Bayesian computing with {INLA}: {N}ew features.
\newblock {\em Computational Statistics \& Data Analysis}, 67:68--83, 2013.

\bibitem{breslow1993}
D.~G.~Clayton N.~E.~Breslow.
\newblock Approximate inference in generalized linear mixed models.
\newblock {\em Journal of the American Statistical Association}, 88(421):9--25,
  1993.

\bibitem{art585}
A.~Riebler, S.~H. S{\o}rbye, D.~Simpson, and H.~Rue.
\newblock An intuitive {B}ayesian spatial model for disease mapping that
  accounts for scaling.
\newblock {\em Statistical Methods in Medical Research}, 25(4):1145--1165,
  2016.

\bibitem{book80}
H.~Rue and L.~Held.
\newblock {\em Gaussian {M}arkov Random Fields: {T}heory and Applications},
  volume 104 of {\em Monographs on Statistics and Applied Probability}.
\newblock Chapman \& Hall, London, 2005.

\bibitem{art375}
H.~Rue and S.~Martino.
\newblock Approximate {B}ayesian inference for hierarchical {G}aussian {M}arkov
  random fields models.
\newblock {\em Journal of Statistical Planning and Inference},
  137(10):3177--3192, 2007.
\newblock Special Issue: {B}ayesian Inference for Stochastic Processes.

\bibitem{art451}
H.~Rue, S.~Martino, and N.~Chopin.
\newblock Approximate {B}ayesian inference for latent {G}aussian models using
  integrated nested {L}aplace approximations (with discussion).
\newblock {\em Journal of the Royal Statistical Society, Series B},
  71(2):319--392, 2009.

\bibitem{art632}
H.~Rue, A.~Riebler, S.~H. S{\o}rbye, J.~B. Illian, D.~P. Simpson, and F.~K.
  Lindgren.
\newblock Bayesian computing with {INLA}: {A} review.
\newblock {\em Annual Reviews of Statistics and Its Applications},
  4(March):395--421, 2017.

\bibitem{art631}
D.~P. Simpson, H.~Rue, A.~Riebler, T.~G. Martins, and S.~H. S{\o}rbye.
\newblock Penalising model component complexity: A principled, practical
  approach to constructing priors (with discussion).
\newblock {\em Statistical Science}, 32(1):1--28, 2017.

\bibitem{art521}
S.~H. S{\o}rbye and H.~Rue.
\newblock Scaling intrinsic {G}aussian {M}arkov random field priors in spatial
  modelling.
\newblock {\em Spatial Statistics}, 8(3):39--51, 2014.

\bibitem{geobugs}
D.~J. Spiegelhalter, A.~Thomas, N.~Best, and D.~Lunn.
\newblock {\em WinBUGS User Manual, Version 1.4}, 2002.

\bibitem{art430}
J.~Wakefield.
\newblock Disease mapping and spatial regression with count data.
\newblock {\em Biostatistics}, 8(2):158--183, 2007.

\bibitem{Wakefield2007}
Jon Wakefield.
\newblock Disease mapping and spatial regression with count data.
\newblock {\em Biostatistics}, 8(2):158--183, 2007.

\end{thebibliography}

\end{document}